\documentclass[pre,twocolumn,showpacs,showkeys,nofootinbib,longbibliography,amsmath,amssymb,floatfix]{revtex4-1}

\usepackage{graphicx}
\usepackage{bm}
\usepackage{amsmath}
\usepackage{amssymb}
\usepackage{color}

\newcommand{\corr}{}
\newcommand{\one}{\mathcal{I}}

\def\br{\mathbf{r}}
\def\bk{\mathbf{k}}
\def\bu{\mathbf{u}}
\def\bv{\mathbf{v}}
\def\be{\mathbf{e}}
\def\lhyd{L_{\rm hydro}}

\def\epl{Europhys. Lett.}
\def\pnas{Proc. Nat. Acad. Sci. USA}

\newcommand{\eq}[1]{Eq.~(\ref{#1})}

\newcommand{\rcite}[1]{Ref.~\cite{#1}}

\begin{document}

\title{Theory of anomalous collective diffusion in colloidal monolayers on a
  spherical interface}

\author{Alvaro Dom\'\i nguez} \email{dominguez@us.es}
\affiliation{F\'\i sica At\'omica, Molecular y Nuclear, Universidad de Sevilla, Apdo.~1065,
  41080 Sevilla, Spain}

\date{9 November 2017}

\begin{abstract}

  A planar colloidal monolayer exhibits anomalous collective diffusion
  due to the hydrodynamic interactions. We investigate how this
  behavior is affected by the curvature of the monolayer when it
  resides on the interface of a spherical droplet. It is found that
  the characteristic times of the dynamics still exhibit the same
  anomalous scaling as in the planar case. The spatial distribution,
  however, shows a difference due to the relevance of the radius of
  the droplet. Since \corr{for the droplet} this is both a global
  magnitude, i.e., pertaining the spatial extent of the spherical
  surface, and a local one, i.e., the radius of curvature, the
  question remains open as to which of these two features actually
  dominates \corr{in the case of a generically curved interface}.
  
\end{abstract}



\maketitle

\section{Introduction}

The hydrodynamic interactions between the particles of a colloid,
which are mediated by flows in the embedding ambient fluid, are very
relevant for the dynamics of the colloid, see, e.g.,
\rcite{Dhon96}. \corr{The presence of near boundaries, like an
  interface, affect these interactions, and additionally introduce a
  new player with which the particles interact hydrodynamically. The
  theoretical study of these effects has a long history, see, e.g.,
  Refs.~\cite{JFD75,AdBl78,LeLe80} for the case of a planar interface
  between two coexisting fluids.
  More recently, one has considered the case when the interface has a
  richer rheological behavior, namely surface viscosity
  \cite{SaDe75,DADL95,BCL99,Feld06c}, elasticity \cite{Feld06b},
  ultra--low surface tension \cite{Bick07}, bending rigidity
  \cite{Bick06,Voro08}, or when it is curved
  \cite{FKJ89,DLG17,ShAr17}.  All these works study the case of a
  single particle and are primarily concerned with
  \emph{self--diffusion}, i.e., the random motion of a tagged
  particle. Our goal is, however, the \emph{collective diffusion},
  that describes the decay of density perturbations. This is an
  intrinsically many--body problem, for which the hydrodynamic
  interaction \emph{between} the particles (but modified by the
  presence of the interface) is most relevant. These are two distinct,
  albeit related concepts\footnote{\corr{For instance, a collection of
      independent, noninteracting particles (ideal gas) is a physical
      realization of an ensemble of isolated particles, so that the
      coefficient of collective diffusion coincides trivially with the
      coefficient of self--diffusion.}}  \cite{Dhon96}. }

Most works addressing the influence of the hydrodynamic interactions
on the collective diffusion have dealt with the case of colloids in
bulk, i.e., three-dimensional (3D) distributions of particles
\cite{Batc76,CiFe91,VdC99}.  \corr{Recent investigations have
  considered confined configurations \cite{Diam09}, e.g.,
  two-dimensional (2D) distributions inside a fluid also confined to
  2D, either between plates \cite{PeNa00a,CDLR04,CLA04} or as a film
  \cite{DKIL08}. A particularly interesting case is a colloidal
  monolayer, produced when the particles are constrained to reside on
  a fluid--fluid interface, see, e.g., \rcite{Bink02}.} It is a
\emph{partially} confined system in that the particle distribution is
confined to a 2D manifold, but the ambient fluid is unconfined in
3D. Recent theoretical investigations, confirmed experimentally
\cite{LCXZ14,BDO15}, predicted that both the short--time \cite{NKPB02}
and the long--time \cite{BDGH14} coefficient of \emph{collective}
diffusion for a planar monolayer diverge, i.e., \corr{the diffusive
  decay of a density perturbation in the monolayer can be described as
  anomalous} due to the hydrodynamic interactions. This feature is
specific to the configuration of partial confinement and is a direct
consequence of the ``dimensional mismatch'' between the 2D colloidal
subsystem and the 3D embedding fluid (see the discussion after
\eq{eq:lincont}). \corr{Numerical simulations \cite{PPD17} suggest
  that this mismatch does not have, however, any dramatic effect on
  the coefficient of \emph{self}--diffusion, which remains finite.}

\corr{One may wonder how robust the anomalous collective diffusion is,
  and so recent works have explored this phenomenology when the
  simplifying assumptions of the original theoretical model are
  relaxed:} one has considered the influence of the direct
particle--particle interaction, e.g., as capillary monopoles
\cite{BDGH14}, as hard spheres \cite{GNK16}, or as Lennard-Jones
particles \cite{PPD17}. One has also addressed the effect of the
finite time it takes for the ambient flow to respond to the evolution
of the colloidal monolayer \cite{Domi14}, or the possibility, beyond
the perfect confinement to a plane, that the particles move slightly
in and out of the plane \cite{PPD17,BDO17}. Along the line of these
investigations, the present work addresses how the role of the
hydrodynamic interactions is affected when the monolayer is curved
rather than perfectly flat.

The curvature of the interface can affect the diffusive dynamics and
alter Fick's law for Brownian diffusion \corr{qualitatively}
\cite{BrLe78,vanK86,DeMo04}. Even when this change is neglected, the
analytical study of diffusion on a curved manifold poses its own
mathematical problems, which one can try to manage by means of
specific tools from the realm of differential geometry, see, e.g.,
Refs.~\cite{Fara02,Cast14}. For the problem at hand, the issue is
further complicated because the determination of the hydrodynamic
interactions requires solving the hydrodynamic equations for the
ambient flow together with the boundary conditions imposed by a curved
manifold. Thus, in this work we consider the simplest configuration of
a perfectly spherical interface supporting the monolayer. This case is
of actual relevance for the interpretation of experimental results,
since the assembly of a monolayer at the surface of a spherical
droplet is a quite common and relatively easy procedure. Furthermore,
this case is amenable to a mathematical analysis allowing for the
derivation of analytical results. On the minus side, this
configuration is very simple and some questions regarding the
influence of curvature on the hydrodynamic interactions remain
open. In Sec.~\ref{sec:theory}, \corr{we introduce and solve the
  simplest model} that exhibits the phenomenology of interest, namely,
the interplay between the intrinsic dynamics of the colloid and the
hydrodynamic interactions mediated by the ambient fluids. The
discussion of the results and the conclusions are presented in
Sec.~\ref{sec:discussion}.

\section{Theoretical model}
\label{sec:theory}

We consider a collection of colloidal particles trapped at the fluid
interface of a spherical droplet \corr{at rest}. The radius of the
droplet will be denoted by $R$, while $\eta_1$ and $\eta_2$ represent
the dynamic viscosities of the fluids outside and inside of the
droplet, respectively. We take spherical coordinates $(r,\theta,\phi)$
with origin at the center of the droplet, so that $\be_r$ denotes the
unit vector normal to the particle monolayer dwelling on the fluid
interface; consequently, the dyadic $\one - \be_r\be_r$ denotes the
projector onto the plane tangent to it (with $\one$ the unit tensor),
and
\begin{equation}
  \label{eq:1}
  \nabla_\parallel := \left. (\one - \be_r\be_r)\cdot\nabla \right|_{r=R}
  = \frac{\be_\theta}{R} \frac{\partial}{\partial\theta} + 
  \frac{\be_\phi}{R \sin\theta} \frac{\partial}{\partial\phi} ,
\end{equation}
is the nabla operator on the spherical surface.

The areal number density of particles in the monolayer is described by
the field $\varrho(\br = R \be_r(\theta,\phi),t)$ defined on the
spherical interface. \corr{It obeys the continuity equation on static
  curved surfaces \cite{Ston90}},
\begin{equation}
  \label{eq:cont}
  \frac{\partial \varrho}{\partial t} = - \nabla_\parallel\cdot (\varrho \bv_\parallel) .
\end{equation}
Here $\bv_\parallel$ is the velocity field of the monolayer, defined
likewise on the spherical interface and tangential to it. \corr{We
  restrict ourselves to long time scales such that the overdamped
  approximation holds \cite{Dhon96}. The flow of the monolayer is
  driven by the gradient of the chemical potential $\mu(\varrho)$ (the
  ``thermodynamic'' force) \cite{Batc76,BDO15}, and by the drag by the
  ambient flow $\bu(\br)$ induced in the surrounding fluids,
  \begin{equation}
    \label{eq:v}
    \bv_\parallel = - \Gamma \nabla_\parallel \mu + \bu(\br\in\mathrm{monolayer}) ,
  \end{equation}
  where $\Gamma$ is the mobility. With the ideal gas approximation,
  \begin{equation}
    \label{eq:mu}
    \mu=-kT\,\ln\varrho 
  \end{equation}
  (here, $k$ is Bolztmann's constant and $T$ is the temperature of the
  system), the first term in \eq{eq:v} yields Fick's law of Brownian
  diffusion on the interface with the surface
  diffusivity 
  $D=\Gamma k T$ \cite{BrLe78}.} (Notice that, because the spherical
interface is assumed impenetrable, the component of the ambient flow
$\bu$ normal to it vanishes, see \eq{eq:unormal} below, so that the
field $\bv_\parallel$ constructed according to this prescription is
indeed tangential).

To provide a complete model, the ambient flow $\bu(\br)$ driven by the
dynamics in the monolayer has to be determined. Unlike the monolayer
fields $\varrho(\br=R\be_r)$ and $\bv_\parallel(\br=R\be_r)$, the
field $\bu(\br)$ is defined everywhere in space. \corr{For colloids,
  it is a good approximation \cite{Dhon96} to use the Stokes
equations describing creeping flow (small Reynolds and Mach numbers)},
\begin{equation}
  \label{eq:u}
  \eta \nabla^2 \bu - \nabla p = 0 ,
  \qquad
  \nabla\cdot\bu=0 ,
\end{equation}
where $p$ is the pressure field enforcing the incompressibility
constraint, and the viscosity $\eta$ takes the value $\eta_1$ or
$\eta_2$, depending on where the equations are considered, i.e.,
outside or inside of the spherical interface. These equations have to
be complemented by the appropriate boundary conditions. Thus, the
velocity is assumed to vanish at infinity \corr{(i.e., no externally
  driven flows)},
\begin{equation}
  \label{eq:uinfty}
  \bu(\br) \to 0 
  \quad
  \mathrm{as}
  \quad
  |\br|\to\infty,
\end{equation}
while, at the interface $\br=R\be_r$,  the normal component
of the velocity vanishes (impenetrable interface),
\begin{equation}
  \label{eq:unormal}
  \be_r\cdot\bu (\br=R^-\be_r) = \be_r\cdot\bu (\br=R^+\be_r) = 0,
\end{equation}
the tangential component is continuous,
\begin{equation}
  \label{eq:utan}
  (\one-\be_r\be_r)\cdot\bu(\br=R^+ \be_r)
  = (\one-\be_r\be_r)\cdot\bu(\br=R^- \be_r) ,
\end{equation}
and the viscous stress
$\sigma := \eta [ \nabla\bu + (\nabla\bu)^\dagger ]$ has a
discontinuity \corr{in the tangential component},
\begin{equation}
  \label{eq:stress}
  (\one-\be_r\be_r) \cdot \left\{ 
    \sigma(\br=R^+ \be_r) - \sigma(\br=R^- \be_r)
  \right\}
  \cdot \be_r 
  = \varrho\nabla_\parallel\mu .
\end{equation}
\corr{This expresses a force balance condition, like \eq{eq:u} but
  localized at the interface. It describes the shear flow driven by
  the Brownian motion in the monolayer. A boundary condition on the
  normal component of the stress is not necessary to solve the
  problem; it only plays a role in order to determine the local forces
  necessary to maintain the surface of the droplet undeformed in spite
  of the presence of the particles and the ambient fluid. In real
  experiments, this constraint is usually achieved by the surface
  tension due to its large value in typical
  interfaces\footnote{\corr{See, e.g.,
      Ref.~\cite[][Suppl.Mat.]{DMPD16a} for a detailed discussion of
      the case of small capillary number.}}.}

\corr{The model just presented provides a coarse-grained description
  of the large scale evolution of the particle distribution. It
  includes implicitly the microscopic details pertaining the shape and
  size of the particles as well as their interactions --- with each
  other and with the fluids and the interface. But it considers
  both the simplest intrinsic dynamics of the colloid (free Brownian
  motion) and the simplest form of hydrodynamic interactions
  (macroscopic drag), which come in with a number of simplifications
  \cite{BDO15}. First, the model leaves complex rheological properties
  of the interface out of consideration; it behaves simply as a
  passive constraint on the particles forcing them to remain attached
  to it. Second, the model also neglects possible modifications of
  Fick's law altogether due to the curvature of the monolayer. Third,
  the direct interaction between the particles --- electric and
  dispersion forces, hard--core effects, etc. are disregarded. The
  only possible interparticle forces are transmitted by the ambient
  fluid, and this hydrodynamic interaction is finally modeled in the
  point--particle approximation: each particle is passively dragged
  (see \eq{eq:v}) by the ambient flow $\bu$ created by the force
  acting on the monolayer (see \eq{eq:stress}), an approach valid for
  a sufficiently dilute monolayer and which can be actually termed
  mean--field--like\footnote{\corr{More precisely, the ambient flow is
      the superposition of the velocity fields created by the force
      acting on each particle \emph{as if} isolated, and each one of
      them experiences this flow \emph{as if} it were created by
      distant sources, e.g., like an externally imposed flow. See,
      e.g., Ref.~\cite[][App.A]{BDO15}
      and 
      Ref.~\cite[][Suppl.Mat.]{BDGH14} for a more detailed
      discussion.}}.}

\corr{All these approximations could be relaxed at the expense of
  mathematical simplicity. Rheological properties of the interface can
  be incorporated in different ways; for instance, surface viscosity
  would appear as an additional term (Boussinesq--Scriven) in
  \eq{eq:stress}. The interfacial curvature can alter Fick's law in
  several ways: from a simple renormalization of the diffusion
  coefficient (e.g., by thermally activated fluctuations in the
  interfacial curvature \cite{ReSe05}) to a scale--dependent diffusion
  coefficient (e.g., by changes in the local curvature on the
  microscopic scale of the monolayer \cite{DeMo04}). In the extreme
  case, even the form of Fick's law could cease to be valid, with
  changes depending on the precise microscopic physics ruling the
  system \cite{vanK86}. The direct interactions are negligible in the
  dilute limit but they can be easily incorporated into the model
  through the density dependence of the chemical potential
  $\mu(\varrho)$ in \eq{eq:v}. This shows up eventually as a
  density--dependent diffusion coefficient, which however does not
  affect the anomalous diffusion phenomenology described by the
  linearized equation~(\ref{eq:lincont}) below. Similarly,
  short--distance corrections to the hydrodynamic interaction due to
  near--neighbours could be incorporated as a density--dependent
  renormalization of the value of the model rheological parameters,
  like the mobility $\Gamma$ \cite{Nozi87,Feld88}.}

\subsection{Linearization}

Equations~(\ref{eq:cont}--\ref{eq:stress}) determine completely the
evolution of the particle number density $\varrho$ in the surface of
the droplet. In order to proceed further, let us assume small
deviations from a homogeneous state,
$\varrho(\br) = \varrho_0 + \delta \varrho(\br)$ with
$|\delta\varrho|\to 0$, and linearize \eq{eq:cont} (all the other
equations are already linear):
\begin{equation}
  \label{eq:lincont}
  \frac{\partial\delta\varrho}{\partial t} \approx 
  D \nabla_\parallel^2 \delta\varrho - \varrho_0 \nabla_\parallel\cdot\bu ,
\end{equation}
This equation still captures the effect both of diffusion by Brownian
motion and of the hydrodynamic interactions between different parts of
the monolayer. Notice that, although $\bu(\br)$ as a 3D field
represents an incompressible flow, see \eq{eq:u}, its restriction to
the 2D monolayer will be compressible in general, so that
$\nabla_\parallel\cdot\bu(\br\in\mathrm{monolayer}) \neq 0$. Together
with the long-range decay of the velocity field given by \eq{eq:u},
this ``dimensional mismatch'' is the ultimate origin of the anomalous
diffusion.

The departure from previous works dealing with this physical problem
is that the monolayer is now a curved manifold. In this particular
case, the mathematical problem can be addressed by expanding the
fields defined on the spherical surface in spherical harmonics
$Y_\ell^m(\theta,\phi)$ (see App.~\ref{app:Y}; the superscript $*$
denotes complex conjugation):
\begin{equation}
  \label{eq:rholm}
  \rho_\ell^m := \int_0^\pi d\theta\; \sin\theta \int_0^{2\pi}d\phi\; Y_\ell^{m*}
  (\theta,\phi)\, \delta\varrho(\theta,\phi) .
\end{equation}
Therefore, equations~(\ref{eq:u}--\ref{eq:lincont}) lead to (see
App.~\ref{app:Y})
\begin{equation}
  \label{eq:lincontlm}
  \frac{\partial\rho_\ell^m}{\partial t} = - D_\ell \frac{\ell
    (\ell+1)}{R^2} \rho_\ell^m ,
\end{equation}
with an effective, $\ell$--dependent diffusion coefficient
\begin{equation}
  \label{eq:Dell}
  D_\ell := D \left[ 1 + \frac{R}{(\ell+1/2) \lhyd} \right] ,
\end{equation}
expressed in terms of the characteristic length
\begin{equation}
  \label{eq:lhyd}
  \lhyd := \frac{4 \eta_+ D}{k T \varrho_0} ,
\end{equation}
which was introduced in Ref.~\cite{BDGH14}, where
$\eta_+ := (\eta_1+\eta_2)/2$ is the average viscosity. (See
App.~\ref{app:flat} for a comparison with the equation for a planar
monolayer).
The solution of \eq{eq:lincontlm} is straightforward,
\begin{equation}
  \label{eq:solrholm}
  \rho_\ell^m(t) =\rho_\ell^m(0) \; \mathrm{e}^{-t/\tau_\ell} ,
\end{equation}
where we have defined the time scales
\begin{equation}
  \label{eq:tauell}
  \tau_\ell := \frac{R^2}{\ell (\ell+1) D_\ell} 
  = \tau_\ell^\mathrm{(norm)} \left[ 1 + \frac{R}{(\ell+1/2) \lhyd}
  \right]^{-1} ,
\end{equation}
\begin{equation}
  \label{eq:taunorm}
  \tau_\ell^\mathrm{(norm)} := \frac{R^2}{\ell (\ell+1) D} .
\end{equation}
In the absence of hydrodynamic interactions, i.e., normal diffusion,
it would be $\tau_\ell = \tau_\ell^\mathrm{(norm)}$ (notice that
$R^2/D$ is the characteristic time for Brownian motion over the size
of the spherical surface).

\begin{figure}[t]
  \centering

  \includegraphics[width=.4\textwidth]{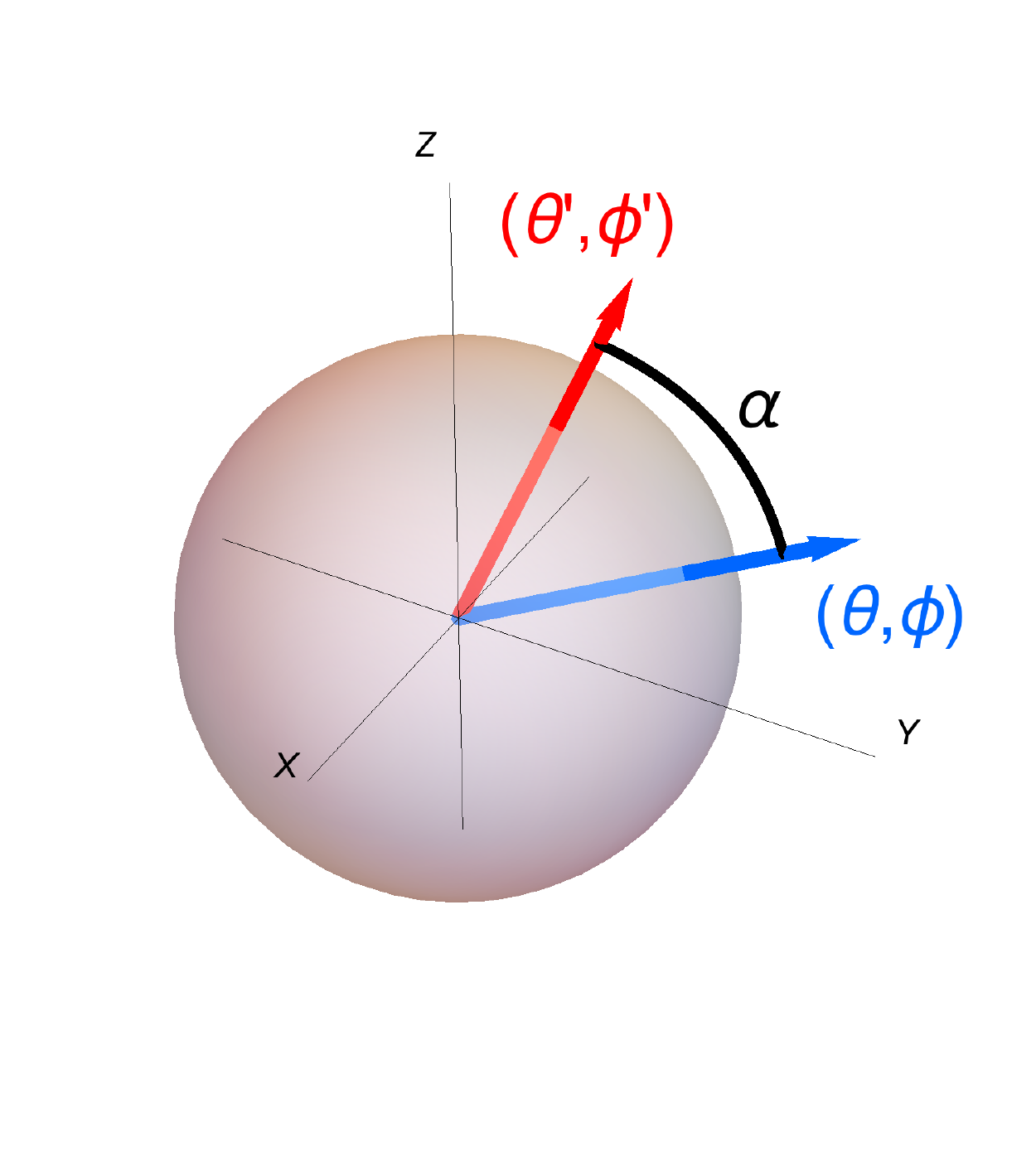}

  \caption{Definition of the angle $\alpha$ used in \eq{eq:green}.}
  \label{fig:alpha}
\end{figure}

The Green function $G$ of \eq{eq:lincont} is defined by the
relationship
\begin{equation}
  \label{eq:G}
  \delta\varrho(\theta,\phi,t) = \int_0^\pi d\theta'\;\sin\theta' \int_0^{2\pi}
  d\phi'\; \delta\varrho(\theta',\phi',0) \; G(\theta,\phi; \theta',\phi'; t) .
\end{equation}
From the solution~(\ref{eq:solrholm}), one can obtain
(see App.~\ref{app:green})
\begin{equation}
  \label{eq:green}
  G(\theta,\phi; \theta',\phi'; t) = \sum_{\ell=0}^\infty
  \frac{2\ell +1}{4\pi} P_\ell (\cos\alpha) \mathrm{e}^{-t/\tau_\ell} ,
\end{equation}
where $\alpha$ is the angle between the directions given by the pairs
$(\theta,\phi)$ and $(\theta',\phi')$, see Fig.~\ref{fig:alpha}.

\section{Discussion} 
\label{sec:discussion}

\begin{figure}[t]
  \centering

  \includegraphics[width=.4\textwidth]{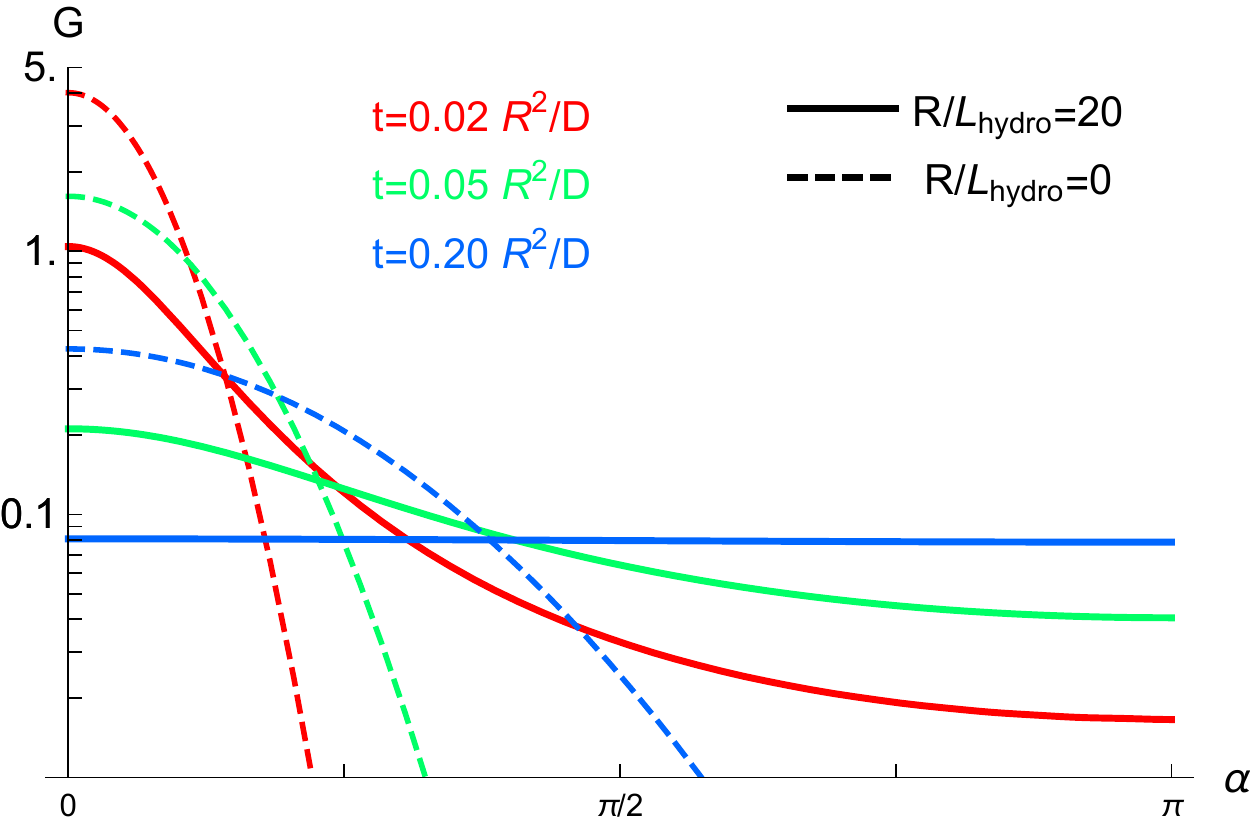}

  \caption{Plot of the Green function, \eq{eq:green}, at different times when the
    hydrodynamic interactions are considered (thick lines)
    or not (dashed lines). 
    The vertical axis is in logarithmic scale.}
  \label{fig:green}
\end{figure}

The effect of the hydrodynamic interactions is already patent in a
comparative plot of the Green function, which formally represents the
diffusion of an initially concentrated distribution,
$\delta\varrho(\theta,\phi,t=0) = (\sin\theta)^{-1} \delta(\theta)
\delta(\phi)$, see Fig.~\ref{fig:green}. Qualitatively, one observes
that the decay in time toward the equilibrium, homogeneous distribution is
faster and the spread in space is broader when the hydrodynamic
interaction is accounted for.

To be more precise, in the limit $R\ll\lhyd$, the time scale defined
by \eq{eq:tauell} behaves as
$\tau_\ell \approx \tau_\ell^\mathrm{(norm)}$ for any value of $\ell$,
so that the effect of the hydrodynamic interactions is
unnoticeable. In the opposite limit $R\gg\lhyd$, however, it is
\begin{equation}
  \frac{\tau_\ell}{\tau_\ell^\mathrm{norm}} \approx \left( \ell + \frac{1}{2}
  \right) \frac{\lhyd}{R} ,
\end{equation}
so that the characteristic times are drastically reduced for the many
large--scale modes satisfying $\ell \lesssim
R/\lhyd$. This ``acceleration'' of the dynamical evolution induced by
the hydrodynamic interactions is a feature shared with the
phenomenology in a planar monolayer; the scaling
$\tau_\ell \sim 1/\ell$, rather than $\tau_\ell \sim 1/\ell^2$ (see
Eqs.~(\ref{eq:tauell}, \ref{eq:taunorm})) justifies the denomination
of ``anomalous diffusion'' \corr{(``superdiffusion'', to be more
  precise)}. Also common is the meaning of the scale $\lhyd$ as a
crossover length for the observation of anomalous diffusion.

Differences arise, however, between both cases (planar and spherical
monolayer) regarding the spatial structure. A useful diagnostic tool
is the average of the Legendre polynomials,
\begin{equation}
  \langle P_\ell(\cos\theta) \rangle = \int_0^\pi d\theta\;\sin\theta
  \int_0^{2\pi}d\phi\;  P_\ell(\cos\theta) \, G(\theta,\phi;0,0,t) ,
\end{equation}
which provide a measure of how the density distribution initially
concentrated at the pole of the sphere spreads over its surface. By
using the orthonormality properties of the Legendre polynomials, it
follows from \eq{eq:green} that
\begin{equation}
  \label{eq:averageP}
  \langle P_\ell(\cos\theta) \rangle = \mathrm{e}^{-t/\tau_\ell} .
\end{equation}
Particularly interesting is the quantity \cite{Cast14}
\begin{eqnarray}
  \label{eq:msd}
  \langle R^2 \sin^2 \theta \rangle 
  & = 
  & \frac{2}{3} R^2 \left[ \langle P_0
    (\cos\theta) \rangle - \langle P_2 (\cos\theta) \rangle \right] 
    \nonumber \\
  & = 
  & \frac{2}{3} R^2
  \left[ 1 - \mathrm{e}^{-t/\tau_2} \right] ,
\end{eqnarray}
\corr{closely related to the second moment of the density
  distribution. It provides a measurement of the lateral extension of}
the diffusing cloud ($R\sin\theta$ is the size projected onto the
equatorial plane $\theta=\pi/2$). \corr{In the case of \emph{normal}
  diffusion in the \emph{plane}, the second moment grows linearly in
  time. This is at variance with the behavior when the hydrodynamic
  interactions are considered:} for an \emph{unbounded} planar
monolayer, the Green function exhibits a tail $\propto r^{-3}$ with
in-plane distance $r$ regardless of the value of the characteristic
length $\lhyd$ (see App.~\ref{app:flat}). This is ultimately a
consequence of the long--ranged nature of the induced ambient flow
\corr{and implies that the average $\langle r^2 \rangle$ is} formally
infinite. To make sense of this magnitude requires a regularization by
means of a large--distance cutoff, e.g., as a finite size of the
system or by relaxing the assumption of instantaneous build--up of the
hydrodynamic interactions \cite{Domi14,BDO16}. \corr{This behavior is
  altered significantly, however, when the interface is spherical. In
  order to obtain a meaningful comparison, consider the short time
  expansion of \eq{eq:msd}, when the difference between the projected
  extension $R\sin\theta$ of the particle cloud and the ``true''
  (geodesic) extension $r=R \theta$ is expected to be statistically
  irrelevant \cite{Cast14}:}
\begin{equation}
  \langle R^2 \sin^2\theta \rangle \approx \frac{2 R^2 t}{3 \tau_2}
  = 4 D_2 t 
  \qquad (t\to 0) .
\end{equation}
\corr{This average is well defined and actually behaves the same as in
  normal diffusion in a plane. The hydrodynamic interactions only show
  up in that the diffusion coefficient $D_2$ is renormalized, see
  \eq{eq:Dell}}. And so, when $R\ll\lhyd$, the hydrodynamic
interactions are irrelevant, $D_2\approx D$, and any mention to the
radius $R$ drops from the expression~(\ref{eq:msd}). In the opposite
limit $R\gg\lhyd$, the diffusion coefficient does depend on the
parameter $R$: it is much larger, $D_2 = (2 R/5\lhyd) D \gg D$, but
still finite, diverging formally only in the limit
$R\to\infty$. \corr{Since $R$ quantifies both the local curvature of
  the interface and its global extension, there remains the ambiguity
  whether $R\to\infty$ should be better interpreted as either the flat
  interface limit or the unbounded interface limit.}

In summary, the dramatic reduction of the diffusion times on scales
above a certain characteristic length $\lhyd$ observed in a flat
monolayer is preserved for a spherical monolayer. In this sense, the
collective diffusion in the spherical configuration can be also
qualified as anomalous. The radius of the spherical interface enters
as a natural cutoff that renders the \corr{second
  moment}~(\ref{eq:msd}) (and, actually, any other higher--order
moment \corr{of the density distribution}) finite. The spherical
configuration, however, is very particular in that the radius is a
quantity pertaining both the \emph{global} structure of the surface,
namely its finite size, and its \emph{local} curvature, \corr{and it
  is not clear how to disentangle the influence of the respective
  features}. Thus, there still remains unanswered the question about
which feature is actually more determinant: could an unbounded, but
locally curved surface disrupt the effect of the hydrodynamic
interactions that leads to anomalous diffusion?

\begin{acknowledgments}
  The author acknowledges support by the Spanish Government through
  Grant FIS2017-87117-P (partially financed by FEDER funds).
\end{acknowledgments}

\appendix

\section{Spherical harmonics}
\label{app:Y}

We use the standard definition of the spherical harmonics,
\begin{equation}
  \label{eq:Y}
  Y_\ell^m(\theta,\phi) := \sqrt{\frac{2\ell+1}{4\pi}
    \frac{(\ell-m)!}{(\ell+m)!}}
  P_\ell^{|m|} (\cos\theta) \mathrm{e}^{i m \phi} ,
\end{equation}
in terms of the associated Legendre functions of the first kind,
$P_\ell^{|m|}$, with $\ell$ a positive integer and $m$ an integer such
that $|m|\leq \ell$. These functions are a complete, orthonormal basis
for functions defined on the surface of a sphere and verify
\begin{equation}
  \label{eq:laplY}
  \nabla_\parallel^2 Y_\ell^m = - \frac{\ell (\ell+1)}{R^2} Y_\ell^m .
\end{equation}

The linear boundary--value problem given by
Eqs.~(\ref{eq:u}--\ref{eq:stress}) can be solved easily with the help
of the spherical harmonics. This is precisely the same problem studied
recently in \rcite{ScSt16}: our Eqs.~(\ref{eq:u}--\ref{eq:stress})
become equations~(1-4) of \rcite{ScSt16} upon identifying
$\nabla_s \sigma \leftrightarrow
\varrho\nabla_\parallel\mu$. \corr{The solution to \eq{eq:u} can be
  written as an expansion in spherical harmonics, with different
  expansion coefficients inside and outside of the spherical
  interface. The boundary
  conditions~(\ref{eq:unormal}--\ref{eq:stress}) at the interface
  provide relationships between the coefficients inside and
  outside. Finally, the boundary condition~(\ref{eq:uinfty}) and the
  additional condition that the velocity field must be regular
  everywhere (in particular, at the origin $r=0$ of the coordinate
  system) determine the value of these coefficients uniquely.} We only
need the velocity field evaluated at points of the monolayer, which is
given by Eq.~(12) in \rcite{ScSt16}; in our notation, it is
\begin{eqnarray}
  \label{eq:umonoY}
  \bu(\br=R\be_r(\theta,\phi)) 
  & = 
  & \mbox{} - \frac{k T R}{2\eta_+}
    \sum_{\ell=1}^\infty \sum_{m=-\ell}^\ell \frac{\rho_\ell^m}{2\ell+1}
    \nabla_\parallel Y_\ell^m (\theta,\phi) ,
\end{eqnarray}
in terms of the average viscosity $\eta_+ := (\eta_1+\eta_2)/2$. The
use of \eq{eq:laplY} renders expression
$\nabla_\parallel\cdot\bu(\br=R\be_r)$ in \eq{eq:lincont}
into an expansion in spherical harmonics, from which \eq{eq:lincontlm}
follows straightforwardly.

\section{The planar monolayer}
\label{app:flat}

For an unbounded, planar monolayer, one introduces the 2D Fourier
transform of a density perturbation,
\begin{equation}
  \rho(\bk) = \int d^2\br\; \mathrm{e}^{-i\bk\cdot\br}
  \delta\varrho(\br) ,
\end{equation}
where $\br=(x,y)$ is a point of the monolayer plane $z=0$. This
quantity obeys the dynamical equation \cite{BDGH14}
\begin{equation}
  \frac{\partial\rho(\bk)}{\partial t} = - D_k^\mathrm{(flat)} k^2 \rho(\bk) ,
\end{equation}
with the diffusion coefficient
\begin{equation}
  D_k^\mathrm{(flat)} := D \left[ 1 + \frac{1}{\lhyd k} \right] .
\end{equation}
The comparison with Eqs.~(\ref{eq:lincontlm}) and (\ref{eq:Dell})
shows that for the small--scale modes ($\ell\gg 1$), they reduce to
the planar case with the identification $k\leftrightarrow \ell/R$. The
large--scale modes are sensitive to the curvature of the spherical
interface and differences between both cases arise.

An analytic expression for the Green function in the planar case,
defined analogously to \eq{eq:G}, can be obtained in the limit
$r\gg \lhyd$ \cite{BDO15},
\begin{equation}
  G(r,t) \approx \frac{1}{2\pi} \left(\frac{\lhyd}{D t}\right)^2
  \left[
    1 + \left(\frac{r \lhyd}{D t}\right)^2 \right]^{-3/2} .
\end{equation}
As a consequence of the slow $1/r^3$ asymptotic decay, \corr{the second
moment of the Green function,}
\begin{equation}
  \langle r^2 \rangle = \int d^2\br \; r^2 G(r,t),
\end{equation}
is undefined in an unbounded monolayer.

\section{The Green function}
\label{app:green}

\corr{The solution~(\ref{eq:solrholm}) allows one to write the
  time--evolved density field as
  \begin{eqnarray}
    \delta\varrho(\theta,\phi,t) 
    & = 
    & \sum_{\ell=0}^\infty
      \sum_{m=-\ell}^\ell \rho_\ell^m(0) \; \mathrm{e}^{-t/\tau_\ell} 
      Y_\ell^{m}  (\theta,\phi)
    \nonumber \\
    \textrm{\tiny (\eq{eq:rholm})\;}
    & = 
    & \sum_{\ell=0}^\infty
      \sum_{m=-\ell}^\ell \mathrm{e}^{-t/\tau_\ell} 
      Y_\ell^{m}  (\theta,\phi)
    \\
    &
    & \mbox{}\times
      \int_0^\pi d\theta'\; \sin\theta' \int_0^{2\pi}d\phi'\; Y_\ell^{m*}
      (\theta',\phi')\, \delta\varrho(\theta',\phi',0) .
      \nonumber 
  \end{eqnarray}
The comparison with \eq{eq:G} gives } the expression
\begin{equation}
  \label{eq:Glm}
  G(\theta,\phi; \theta',\phi'; t) = \sum_{\ell=0}^\infty
  \sum_{\ell=-m}^m \mathrm{e}^{-t/\tau_\ell} 
  Y_\ell^{m}(\theta,\phi)  Y_\ell^{m*}(\theta',\phi') .
\end{equation}
This can be simplified further by using the definition of the
spherical harmonics, \eq{eq:Y}, and by applying the addition theorem
\cite{GrRy94},
\begin{eqnarray}
  \label{eq:add}
  P_\ell (\cos\alpha) 
  & = 
  & P_\ell (\cos\theta) P_\ell (\cos\theta')
  \\
  & \mbox{} + 
  & 2 \sum_{m=1}^\ell \frac{(\ell-m)!}{(\ell+m)!} P_\ell^m
  (\cos\theta) P_\ell^m (\cos\theta') \cos m(\phi-\phi') ,
    \nonumber 
\end{eqnarray}
where the angle $\alpha$ (see Fig.~\ref{fig:alpha}) satisfies
\begin{equation}
  \label{eq:cosalpha}
  \cos\alpha =\cos\theta \, \cos\theta' +  \sin\theta \, \sin\theta' \cos (\phi-\phi') .
\end{equation}
In this manner, \eq{eq:Glm} is simplified to \eq{eq:green}.


%

\end{document}